\documentstyle[epsf]{article}
\begin{document}
\renewcommand{\thefootnote}{\fnsymbol{footnote}}
\begin{flushright}
KEK-preprint-94-177\\
DPNU-94-58\\
NWU-HEP 94-08\\
TIT-HPE-94-13\\
TUAT-HEP 94-08\\
OCU-HEP 94-10\\
PU-94-691\\
INS-REP 1076\\
KOBE-HEP 94-09\\
\end{flushright}
\vskip -3cm
\epsfysize3cm
\epsfbox{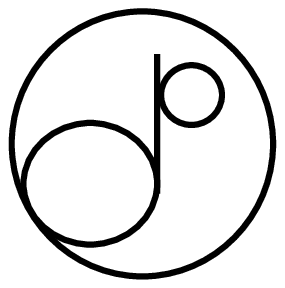}
\begin{center}
{\large \bf
Observation of Excess
$\Lambda(\overline{\Lambda})$ Production in Two-Photon Processes at TRISTAN
\footnote{
submitted for publication.
}
}\\
\vskip 0.5cm
(TOPAZ Collaboration)\\
\vskip 0.5cm
\underline{R.Enomoto}$^a$\footnote{Internet address: enomoto@kekvax.kek.jp.},
K.Abe$^b$, T.Abe$^b$, I.Adachi$^a$,
K.Adachi$^c$, M.Aoki$^d$, M.Aoki$^b$, S.Awa$^c$,
K.Emi$^e$, H.Fujii$^a$, K.Fujii$^a$,T.Fujii$^f$, J.Fujimoto$^a$,
K.Fujita$^g$, N.Fujiwara$^c$, H.Hayashii$^c$,
B.Howell$^h$, N.Iida$^a$, R.Itoh$^a$, Y.Inoue$^g$, H.Iwasaki$^a$,
M.Iwasaki$^c$, K.Kaneyuki$^d$, R.Kajikawa$^b$,
S.Kato$^i$, S.Kawabata$^a$, H.Kichimi$^a$, M.Kobayashi$^a$,
D.Koltick$^h$, I.Levine$^h$, S.Minami$^d$,
K.Miyabayashi$^b$, A.Miyamoto$^a$, K.Muramatsu$^c$, K.Nagai$^j$,
K.Nakabayashi$^b$, E.Nakano$^b$,
O.Nitoh$^e$, S.Noguchi$^c$, A.Ochi$^d$, F.Ochiai$^k$,
N.Ohishi$^b$, Y.Ohnishi$^b$, Y.Ohshima$^d$,
H.Okuno$^i$, T.Okusawa$^g$,T.Shinohara$^e$, A.Sugiyama$^b$,
S.Suzuki$^b$, S.Suzuki$^d$, K.Takahashi$^e$, T.Takahashi$^g$,
T.Tanimori$^d$, T.Tauchi$^a$, Y.Teramoto$^g$, N.Toomi$^c$,
T.Tsukamoto$^a$, O.Tsumura$^e$, S.Uno$^a$, T.Watanabe$^d$,
Y.Watanabe$^d$, A.Yamaguchi$^c$, A.Yamamoto$^a$,and M.Yamauchi$^a$\\
\vskip 0.5cm
{\it
$^a$KEK, National Laboratory for High Energy Physics, Ibaraki-ken 305, Japan \\
$^b$Department of Physics, Nagoya University, Nagoya 464, Japan\\
$^c$Department of Physics, Nara Women's University, Nara 630, Japan \\
$^d$Department of Physics, Tokyo Institute of Technology, Tokyo 152, Japan\\
$^e$Dept. of Appl. Phys., Tokyo Univ. of Agriculture and
Technology, Tokyo 184, Japan\\
$^f$Department of Physics, University of Tokyo, Tokyo 113, Japan\\
$^g$Department of Physics, Osaka City University, Osaka 558, Japan \\
$^h$Department of Physics, Purdue University, West Lafayette, IN 47907, USA \\
$^i$Institute for Nuclear Study, University of Tokyo, Tanashi,
Tokyo 188, Japan \\
$^j$The Graduate School of Science and Technology, Kobe University, Kobe 657,
Japan \\
$^k$Faculty of Liberal Arts, Tezukayama University, Nara 631, Japan \\
}
\end{center}
\begin{abstract}
We have carried out inclusive measurements of
$\Lambda(\overline{\Lambda})$
production in two-photon
processes at TRISTAN. The mean $\sqrt{s}$ was 58 GeV and the
integrated luminosity was 265 pb$^{-1}$.
Inclusive $\Lambda (\overline{\Lambda})$
samples were obtained under such conditions as
no-electron, anti-electron, and remnant-jet tags.
The data were compared with theoretical calculations.
The measured cross sections are two-times larger than the
leading-order theoretical
predictions,
suggesting the necessity of next-to-leading-order
Monte-Carlo generator.
\end{abstract}
\section{Introduction}

Charm pair production in two-photon processes
at $\sqrt{s}$=58 GeV have been reported
in references \cite{exclusive,inclusive,ks,electron,uehara}.
All of these papers have reported larger cross sections than
that based on the
lowest-order (LO) two-photon theory \cite{direct,resolved},
when the Drees-Grassie
parametrization \cite{dg} in the resolved-photon processes
was taken into account.
In our previous reports \cite{inclusive,ks,electron} we
showed that the charm-pair cross sections
approximately agreed with theory when we took into account the
next-to-leading-order correction (NLO) to the hard interactions
\cite{drees},
the lower charm-quark mass, and the intrinsic
gluon $P_T$ inside a photon
\cite{pythia}, together with the Levy-Abramowicz-Charchula set-1 (LAC1)
parametrization
\cite{lac1}
for the resolved-photon process.
This explanation of excesses, however, must be confirmed by
further analysis.

In this article we present an analysis of inclusive
$\Lambda(\overline{\Lambda})$ production in the two-photon
reaction using
data taken with the TOPAZ detector at TRISTAN.
As was demonstrated in our $K_s$ analysis \cite{ks},
strange particles enhance the charm fraction
in a sample.
This analysis is, thus, expected to provide another
handle for checking the
above-mentioned theoretical arguments.
Moreover, since
$\Lambda(\overline{\Lambda})$ is, being a strange baryon,
favorably produced
from gluon jets \cite{diquark,lundmodel},
we expect to observe the NLO effects directly by measuring
the inclusive $\Lambda (\overline{\Lambda})$ production rates.

In addition, the technique used to tag a remnant-jet
using a forward calorimeter
\cite{fcl}, which was developed in the $K_s$ analysis
\cite{ks}, can also be applied in this analysis.
We have thus derived inclusive $\Lambda(\overline{\Lambda})$ cross
sections for both direct and resolved-photon processes
separately.

\section{Event selection}
The data used in this analysis were obtained with the TOPAZ detector
at the TRISTAN $e^+e^-$ collider, KEK\cite{topaz,tpc}.
The mean $\sqrt{s}$ was 58 GeV and the integrated luminosity
was 265 pb$^{-1}$.
A forward calorimeter (FCL), which covered
$0.98<|\cos\theta|<0.998$ ($\theta$ is the
polar angle, i.e., the angle with respect to the electron beam),
was installed during the course of the experiment.
The FCL was made of bismuth germanate
crystals (BGO), and was used to anti-tag the beam electrons (positrons)
and to tag hadrons (remnant-jets) \cite{fcl}.
The integrated luminosity of the data with the FCL
detector was 241 pb$^{-1}$.

A description of our trigger system can be found in
reference \cite{trigger}. The requirement for the charged-track trigger
was two or more tracks with an opening angle greater than
45-90 degrees. The $P_T$ threshold for charged particles was 0.3-0.7 GeV,
which varied depending on the beam conditions.

The event-selection criteria were as follows:
there had to be
three or more charged particles ($P_T>0.15$ GeV, $|\cos\theta|<$0.77),
the invariant mass ($W_{VIS}$) of visible particles
($|\cos\theta|<$0.77) had to be greater than 2 GeV,
the event-vertex position had to be consistent with the interaction point,
and the visible energy
had to be less than 25 GeV.
In total, 280673 events were selected.

\section{Monte-Carlo simulation}
In order to estimate the acceptances and backgrounds in this analysis,
we used the following Monte-Carlo simulation programs.
Details concerning the
event generation of
direct
as well as
resolved-photon
and
vector meson dominance (VDM) processes
can be found in references \cite{exclusive,inclusive,hayashii}.
Here, we just note the following points.
For $c\bar{c}$ generation,
we used the current charm-quark mass of 1.3 GeV
to calculate the cross sections for point-like processes and a
constituent charm-quark mass of 1.6 GeV
for the hadronization procedure, and made a
next-to-leading order (NLO) correction by factorization,
the details of which can be found in
references \cite{exclusive,inclusive,drees}.
Light-quark generation was carried out by using the lowest
order (LO) formula with a $P_T^{min}$ cut of 2.5 GeV.
We used the parton density functions by LAC1
\cite{lac1} for the resolved-photon process,
because the experimental data have favored this parametrization
\cite{inclusive,ks,electron}.
JETSET6.3 \cite{lund} was used for hadroniztion of parton system.
Here, we must mention that we used the default values for the baryon production
parameters, such as a $P(qq)/P(q)$ ratio of 0.10 and a $BM\overline{B}$ ratio
of 0.5.
Generated events were processed through the standard TOPAZ detector
simulation program \cite{adachi}, in which
hadron showers were simulated with an extended version of
GHEISHA 7 \cite{gheisha}.
The simulation's handling of hadronic
interactions with nuclei has been updated to fit various
experimental cross sections.

Using the above-mentioned Monte-Carlo simulations, the trigger
efficiency for the sum of the direct and resolved-photon processes was
estimated to be 79\%, 97\% of which represented charged trigger events.
The event-selection efficiency after the trigger was obtained to
be 80\%.

To generate single-photon-exchange hadronic events we used
JETSET6.3\cite{lund} with the parameter values given in
reference \cite{adachi}.

\section{Tagging conditions}

The tagging conditions were as follows. For
anti-electron tagging there had to be no
energy deposit of more than 0.4$E_b$
in $|\cos\theta|<0.998$ (anti-electron tag or anti-tag),
where $E_b$ is the
beam energy.
The rejected sample was called the electron tag sample.
These selected events were from collisions of almost-real
photons, for which the equivalent photon approximation
was expected to be accurate at the 1\% level.
When the energy cut was lowered, the number of mistakenly rejected
beam remnant hadron (remnant-jets) events became significant,
as predicted by Monte-Carlo simulations.
This implies the possibility of tagging the resolved-photon process
by requiring, for instance, 500 MeV $<E_{FCL}<$ 0.25$E_b$,
where $E_{FCL}$ is the energy deposit in the FCL (remnant-jet tag
or rem-tag).
The yield of the remnant-jet tag events agreed with our Monte-Carlo
simulation within 5\% \cite{ks}.
An event selection without these two tags is hereafter called,
a ``no-electron tag" or a ``no-tag".
Also, data without using the FCL detector were included in this sample.
The fractions of anti-electron and remnant-jet tag events to no-tag events
were obtained to be 97.6 and 47\%, respectively.

In the Monte-Carlo simulations we used the equivalent photon approximation
with the
photon flux expression \cite{flux}
\begin{eqnarray*}
    f_{\gamma/e}( x_{\gamma})& =& \frac{ \alpha_{em}}{2 \pi x_{\gamma}}
                 \left( 1 + ( 1 - x_{\gamma})^{2} \right)
                 \ln \frac{P^{2}_{max}}{P^{2}_{min}}
           \     -  \ \frac{\alpha_{em}}{\pi}
\frac{1-x_{\gamma}}{x_{\gamma}},
\end{eqnarray*}
where
\begin{eqnarray*}
       P^{2}_{min} & = & m_e^{2} \frac{ x_{\gamma}^{2}}{1 - x_{\gamma}}.
\label{eqn:flux}
\end{eqnarray*}
We set the $P^2_{max}$ limit at the smaller of $P_{T,q}^2+m_q^2$
and the anti-tag limit
[$2E_b^2(1-x_{\gamma})(1-\cos\theta_{max})$,
where $x_{\gamma}=0.4,~\theta_{max}
=3.2$ degrees],
where $P_{T,q}$ and $m_q$ are
the transverse momentum and the mass
of a quark, respectively \cite{whit}.

The tagging efficiency of the remnant jet for the resolved-photon
process was estimated to be $76\pm3$\%,
taking into account an estimate of
the FCL noise (described later).
We tried to generate remnant partons
using two techniques:
one was along the beam direction;
the other used a Gaussian distribution of the $P_T$-width (0.44 GeV)
with respect to the
beam axis. These two methods differed in acceptance by only 1\%.
Further, the tagging efficiency of the remnant-jet tag
for the direct process was estimated to be $20\pm2$\%
with the estimated level of FCL noise.


\section{$\Lambda (\overline{\Lambda})$ inclusive analysis}

\begin{figure}
\vskip -2cm
\epsfysize10cm
\epsfbox{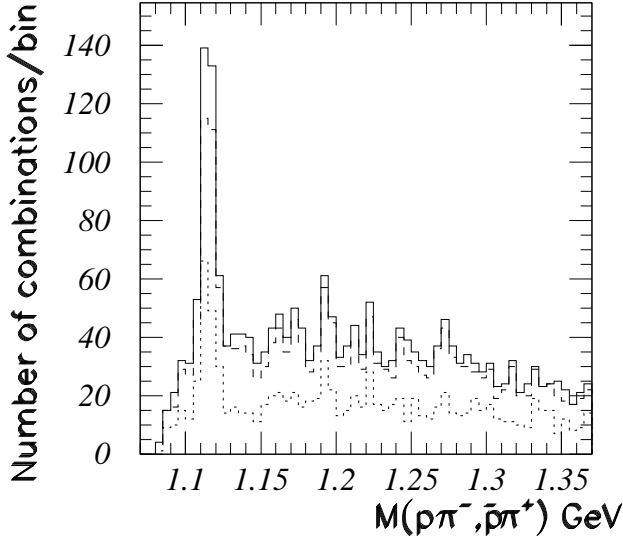}
\vskip -1cm
\caption{
Invariant-mass spectra of $p\pi^-$'s and their charge conjugations.
The solid
histogram is for the no-electron tag, the dashed one is for
the anti-electron tag, and the dotted one is for the
remnant-jet tag.
}
\label{mass}
\end{figure}

The charged-track selection criteria for the
$\Lambda(\overline{\Lambda})$ analysis were as follows:
for each track
$P_T$ had to be greater than 0.15 GeV,
$|\cos\theta|$ had to be less than 0.77, and the closest
approach to the interaction point in the XY-plane (perpendicular
to the beam axis) had to be greater than 0.5 cm.
Using these selected tracks,
we looked for opposite-sign pairs with an opening angle
of less than 90 degrees,
and carried out secondary vertex reconstructions
three-dimensionally.
Here, the
dE/dx of one of two tracks had to be consistent with the proton
assumption ($\chi^2_{p(\overline{p})}<10$), and that of the other
had to be consistent with the pion assumption
($\chi^2_{\pi^{\pm}}<10$).
We demanded that these pairs
be consistent with the assumption that they came
from event vertices with flight lengths greater than 3 cm.
Finally, we rejected pairs which had vertices near
($\pm$3cm) to the inner
pressure vessel ($R_{xy}\sim$30cm) or
the field cage ($R_{xy}\sim$33cm) of the TPC \cite{tpc} in order
to reduce the background from nuclear interactions;
by generating events without
any $\Lambda (\overline{\Lambda})$'s,
we found that there were no significant contribution from such
fake pairs after this cut.
We also required that $|\cos \theta|$ of the pair had to be less than 0.77.
The invariant-mass distributions of these candidate
$p\pi^-$ pairs and their charge-conjugated states (CC)
are plotted in Figure \ref{mass}
for the three tagging conditions, respectively.
These invariant-mass spectra were fitted with the sum of a fifth-order
polynomial and a Gaussian distribution,
and the peak entries were obtained for no-,
anti-, and remnant-jet tags
to be $255\pm20$, $215\pm19$, and $123\pm13$
$\Lambda(\overline{\Lambda})$'s,
respectively, in the $P_T$ range between 0.75 - 2.75 GeV.
The peak position and the width were consistent with the detector
simulation.
Among them, 2-$\Lambda \Lambda$, 0-$\overline{\Lambda} \overline{\Lambda}$,
and 0-$\Lambda \overline{\Lambda}$ pairs were found.
In order to derive the differential $P_T$ cross sections we divided
$P_T$ into four bins, as shown in Table \ref{tcross}.
\begin{table}
\begin{tabular}{cccccc}
\hline
\hline
tag cond. & no-tag & anti-tag & rem-tag & rem-tag
& anti-rem-tag\\
VDM subt. & - & - & yes & no & no \\
\hline
$P_T$ range& \multicolumn{5}{c}{cross sections} \\
 (GeV) & \multicolumn{5}{c}{(pb/GeV)} \\
\hline
 0.75-1
 & 108$\pm$  30& 102$\pm$  30 &  29$\pm$  11&  85$\pm$  28
 &  73$\pm$  28\\
 1-1.25
 &  46$\pm$  12&  42$\pm$  12 &  17$\pm$   6&  35$\pm$  11
 &  25$\pm$  12\\
 1.25-1.5
 &13.2$\pm$ 4.9&12.1$\pm$ 4.7
 & 5.2$\pm$ 3.1& 7.4$\pm$ 4.1
 & 6.9$\pm$ 5.2\\
 1.5-2.75
 & 5.1$\pm$ 1.4& 3.5$\pm$ 1.2
 & 2.3$\pm$ 1.0& 2.4$\pm$ 1.0
 & 1.2$\pm$ 1.4\\
\hline
\hline
\end{tabular}
\caption{
Differential cross section of $\Lambda(\overline{\Lambda})$
versus $P_T$ (GeV)
[$d\sigma/dP_T$ (pb/GeV)], for $|\cos\theta|<0.77$. Six cases
are listed: no-tag , anti-electron tag, remnant-jet tag
with VDM subtraction,
anti-remnant-jet tag without VDM subtraction,
and remnant-jet tag without VDM subtraction,
which are described in the text.
}
\label{tcross}
\end{table}

\section{Background subtractions}

The single-photon-exchange process produced a large background,
especially for high-$P_T$ $\Lambda(\overline{\Lambda})$'s.
This background could have been
reduced by applying a cut on the total visible energy.
However, we avoided this in order
to keep the acceptance for high-$P_T$
$\Lambda(\overline{\Lambda})$'s.
Instead, the contamination from the single-photon-exchange process was
estimated and subtracted using a Monte-Carlo
simulation on a bin-by-bin basis.
The background fractions
for the no-tag sample were $3\pm1$, $5\pm1$, $12\pm3$,
and $16\pm3$\%, respectively,
for the $P_T$ bins shown in Table \ref{tcross};
they were strongly $P_T$ dependent.
We also
estimated these fractions for anti- and remnant-jet tags. They were
consistent with the above-mentioned values within the statistical errors.

The background from beam-gas interactions was estimated using the
off-vertex events in the beam direction.
The beam-gas
contribution
for a no-tag sample was $22\pm3$\% on the average,
and was subtracted from the data.
This was mostly due to a vacuum leak in the beam pipe
during some period.
It was larger than in the case of the $K_s$ analysis \cite{ks}.
For the anti- and remnant-jet tag samples, the above values became
$26\pm4$ and $28\pm5$\%, respectively.
Without the anti-tag condition, the beam-gas
contribution became slightly larger, suggesting that the electron-tagged
events were cleaner.

FCL noise hits were studied by analyzing random-trigger
and Bhabha events.
The probability of noise hits with $E_{FCL}>$0.5 GeV
was estimated to be 13.7\%,
while for hits with $E_{FCL}>0.4E_b$ it reduced to be 0.1\%.
The FCL noise was also related to a vacuum leak.
In the Monte-Carlo simulations we added noise hits randomly
in accordance with the observed noise-hit
probability in order to reliably estimate the tagging efficiencies.

\section{Systematic errors}

The systematic errors
for the cross sections were estimated, bin by bin, as follows.
For the trigger, we added some extra noise hits in the tracking chambers
in the simulations.
For the event selection and the
$\Lambda(\overline{\Lambda})$ reconstruction
we changed the cut values and evaluated
the systematic errors as
cross-section differences.
We also changed the pulse-height threshold in the TPC simulation
in order to evaluate the effects on its tracking efficiency.
We added the obtained systematic errors quadratically on a bin-by-bin
basis.

In order to investigate the effect of nuclear interactions
in the material in front of the TPC, we compared the yields of
$\Lambda$ and $\overline{\Lambda}$ in the experimental data
as well as in the Monte-Carlo data.
The ratio [$N(\Lambda)/N(\overline{\Lambda})$]
was $1.4\pm0.3$ in the experiment, while
the Monte-carlo simulation predicted this value to be $1.43\pm0.09$,
being consistent with the data.
The deviation from 1.0 was considered to be due to an inelastic scattering of
$\overline{\Lambda}$ with the material in front of the TPC.
This occurred mostly in the low-$P_T$ regions.
The effect of nuclear interactions was corrected
using the Monte-Carlo simulations.
The total systematic errors were 16$\sim$21\%, depending on $P_T$,
of which
the cut dependence in the event selection
was the dominant source.
These systematic errors were quadratically added to
the statistical errors.

We also checked the acceptance ambiguity due to the parametrization
dependence
of the resolved-photon processes
by comparing the LAC1 \cite{lac1} and
Drees-Grassie [DG] \cite{dg} parametrizations.
The acceptance difference was estimated to be 3.3\%, which is small
compared to the systematic errors given above.

\section{Results}
\begin{figure}
\vskip -3cm
\epsfysize20cm
\epsfbox{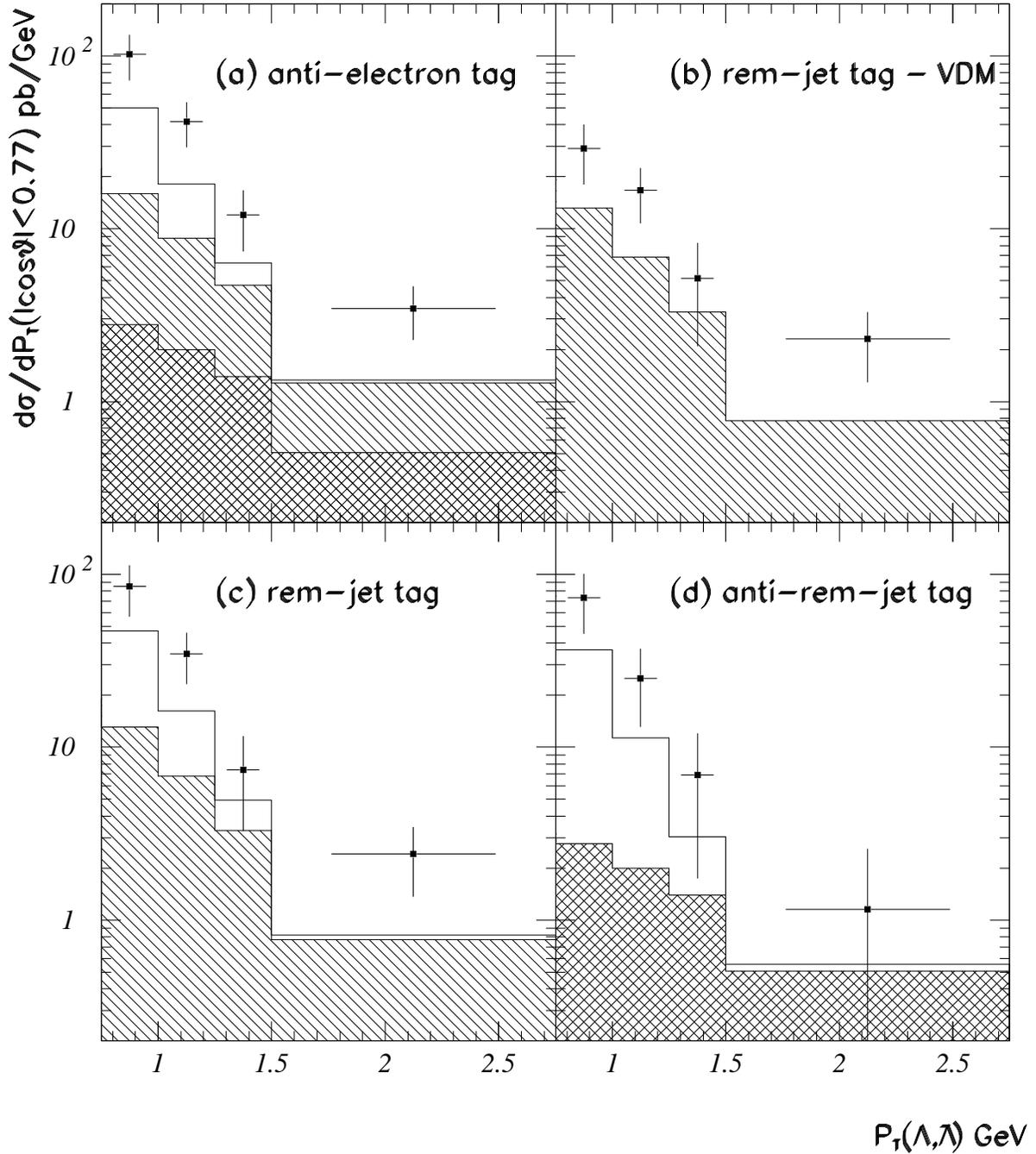}
\caption{
Differential cross section of
$\Lambda(\overline{\Lambda})$ versus $P_T$ (GeV)
[$d\sigma/dP_T$ (pb/GeV)], for $|\cos\theta|<0.77$. Five cases
are plotted: (a) anti-electron tag, (b) remnant-jet tag
with the VDM and direct process subtraction,
(c) remnant-jet tag without the VDM subtraction,
and (d) anti-remnant-jet tag without the VDM subtraction,
as
described in the text.
The corresponding processes are presumably;
(b) resolved-photon, (c) resolved-photon and VDM, and (d) VDM and
direct processes.
The histograms are the theoretical predictions which are described in
the text. The open area is for the VDM, the singly-hatched one is for
the resolved-photon process, and the cross-hatched one is for
the direct process.
}
\label{fcross}
\end{figure}

The $P_T$ differential cross sections were obtained
from the number of reconstructed
$\Lambda(\overline{\Lambda})$'s in each bin, and
its corresponding
efficiency was estimated using the
previously described Monte-Carlo simulations.
They are listed in Table \ref{tcross} and plotted in Figures
\ref{fcross} (a) - (d) for the three tagging conditions
and two subtraction schemes discussed afterward,
respectively.
Figure \ref{fcross} (a) is for anti-electron tag
events.
In the remnant-jet tag events, the Monte-Carlo simulation predicted
a significant contamination from VDM events.
The tagging efficiency for the VDM process was estimated to
be $65\pm6$\%, slightly less than that of the resolved-photon
process. In addition, there was a large ambiguity in the cross
section of the VDM process. We therefore calculated the cross
sections using two subtraction schemes. Figure \ref{fcross}
(b) was obtained by subtracting the VDM contribution
predicted by the Monte-Carlo simulation (VDM subtraction) for
the remnant jet.
Here, the ``anti-remnant-jet tag" cross section was
obtained by subtracting the remnant-jet tag cross section from
that of the anti-electron tag.
We do not show the cross sections for the anti-remnant-jet tag
with VDM subtraction because of low statistics.
Figures \ref{fcross} (c) and (d) were obtained without
VDM subtraction.
In Figures \ref{fcross} (b) and (c), the contribution
of the direct process was subtracted.
The histograms in Figures \ref{fcross} (a) - (d) are the
Monte-Carlo predictions:
the cross-hatched, singly-hatched, and open areas are
predictions for the direct,
resolved-photon (LAC1), and VDM processes, respectively.

\section{Discussions}

The fraction of charm events was studied using the
above-mentioned Monte-Carlo simulations.
We found that 70\% of
the events with
$P_T(\Lambda , \overline{\Lambda})>$ 1.5 GeV were of charm origin.
On the other hand,
only 35\% of the
events with charged tracks of
$P_T>$ 1.5 GeV were from $c\bar{c}$ pairs.
Also, the Monte-Carlo simulations predicted that 55\% of these
high-$P_T$ charm events originated from the direct process.
The contribution of the resolved-photon process is higher than
that in the $K_s$ analysis \cite{ks}.
This is considered to be due to the existence of gluon jets
in the resolved-photon process, even in the LO calculation.
In this study we derived six types of cross sections using
different tagging conditions and subtraction schemes.
We can therefore separately compare
each cross section with the theoretical prediction for each
process.

Firstly, about 30\% of the high-$P_T$ ($P_T>1.5$ GeV) events
can be explained as
electron-tagged events (see Table \ref{tcross}).
Secondly, the anti-tag cross section [Figure \ref{fcross} (a)]
is two-times larger than the theoretical model prediction,
though
the spectrum shape seems to be consistent with it.
Here, we must mention that this theoretical model reasonably
explained the $K_s$ inclusive cross section in the previous
study \cite{ks}.

The cross sections with the remnant-jet
and anti-remnant-jet tags
also show discrepancies having a factor of $\sim$2
compared to
predictions of the direct, resolved, and VDM processes [Figures \ref{fcross}
(b)-(d)].

The VDM model is considered to be ambiguous, especially in
predicting the total
cross sections. However, the $\Lambda (\overline{\Lambda})$-spectrum
of this process is softer than the others
and also than the experimental data.
We therefore cannot fit it to the experimental data by
changing the normalization.
Moreover, increasing the normalization factor makes
the $K_s$ cross-section discrepancy in the low-$P_T$ region
larger, and inconsistent with the experimental
observation \cite{ks}.

\begin{figure}
\vskip -3cm
\epsfysize20cm
\epsfbox{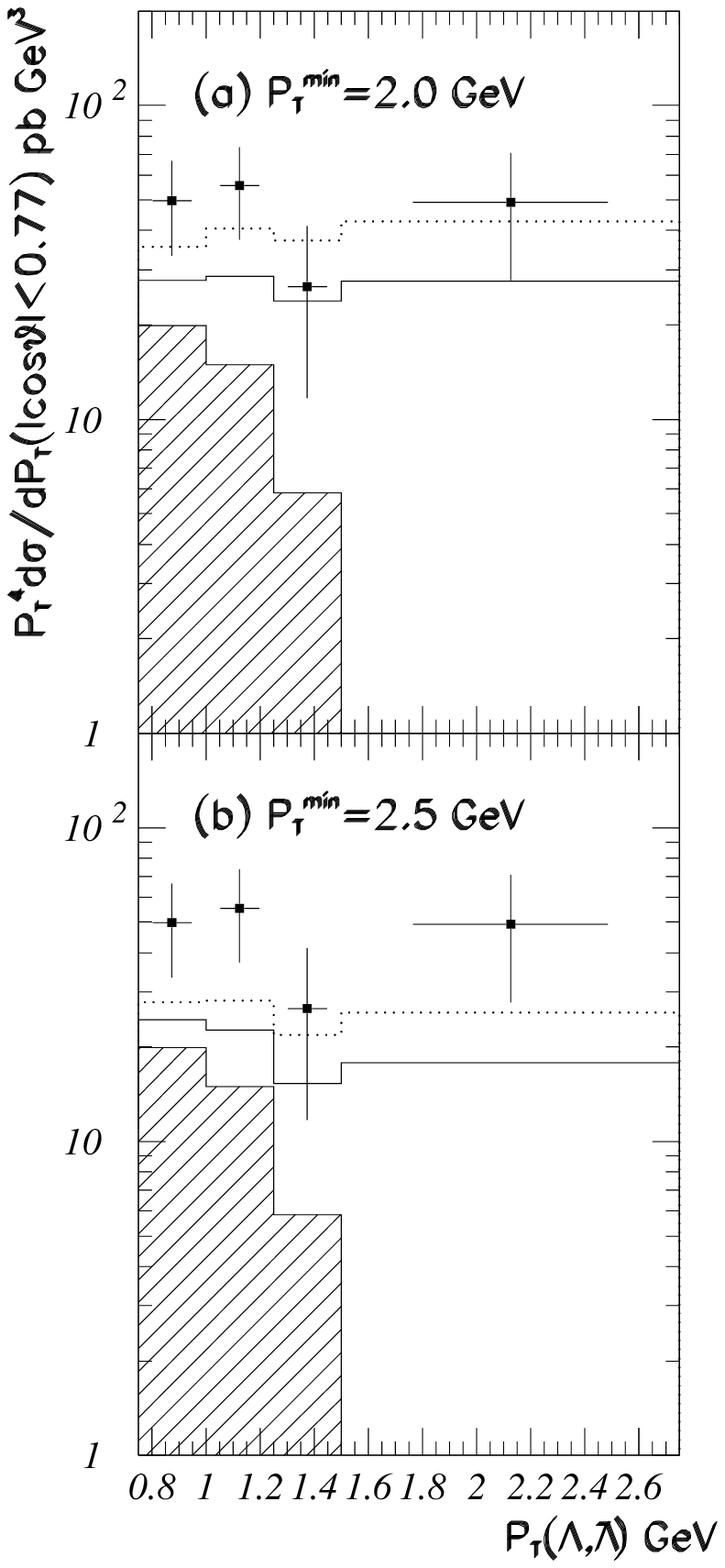}
\caption{
Differential cross section of
$\Lambda(\overline{\Lambda})$ versus $P_T$ (GeV)
[$P_T^4d\sigma/dP_T$ (pb$\cdot$GeV$^3$)],
for $|\cos\theta|<0.77$ for the remnant-jet
tag without the VDM subtraction.
The hatched areas are the predictions by the VDM Monte-Carlo simulation.
The histograms are predictions by the WHIT-1 and -4 parametrizations.
The solid one is WHIT 1
and
the dotted one WHIT-4.
Two values of $P_T^{min}$'s were used, i.e., (a) 2.0 and (b)
2.5 GeV.
}
\label{fres}
\end{figure}

In order to check whether the parton-density functions
have anything to do with the discrepancy,
we compared our remnant-jet-tag data [Figure \ref{fcross} (c)]
with the predictions from six sets of parametrizations by
Hagiwara, Tanaka, Watanabe, and Izubuchi [WHIT1-6] \cite{whit}.
A systematic analysis on the gluon distributions can be carried out
using these parametrizations.
We selected those parametrizations which showed higher charm cross
sections, i.e., WHIT-1 and -4.
The results are shown in Figures \ref{fres} (a) and (b),
where the histograms are the predictions by the WHIT-1 and -4 parametrizations
with $P_T^{min}$'s of 2.0 and 2.5 GeV.
There are some possible combinations
(for example, WHIT-4 with $P_T^{min}=2.0$ GeV seems to fit the
data better).
Notice that these operations do not solve any discrepancies
which appeared in the anti-remnant-jet-tag data.

In order to reduce these discrepancies
we may need an NLO correction to the light-quark events,
which is absent from our
present Monte-Carlo generator.

Our Monte-Carlo generation of $\gamma \gamma$ events was
based on the LO matrix element calculations in the
hard processes.
Although the total cross section of the direct process was increased
by a factor of 1.31 \cite{drees,inclusive,ks,electron}, and
the $P_T$($P_T$ of parton)-dependent factor was used in the
resolved-photon processes \cite{inclusive,ks,electron},
our Monte-Carlo generator has no explicit hard gluon emissions,
which should exist in a truly NLO generator.
We consider that the discrepancies
between the data and the theory can be attributed to
these effects, based on the baryon abundance in the
experimentally observed gluon
jets \cite{diquark,lundmodel}.
The baryon excess in the gluon jets has not yet
been established in a sufficiently quantitative way.
When it is,
our data will be direct evidence of the NLO effect.

We show in Table \ref{ttotal}
the total cross sections of $\Lambda (\overline{\Lambda})$
in the $|\cos \theta |<0.77$ and $0.75<P_T<2.75$ GeV range.
\begin{table}
\begin{tabular}{ccccc}
\hline
\hline
tag cond. & Experiment & Theory (LO) & Exp./Theory & subprocess \\
\hline
antitag & 43.3$\pm$8.3 & 19.1 & 2.26$\pm$0.43 & VDM+resolved+direct \\
rem-tag (-VDM) & 15.6$\pm$3.5 & 6.0 & 2.60$\pm$0.58 & resolved \\
rem-tag & 34.8$\pm$7.8 & 17.3 & 2.01$\pm$0.45 & VDM+resolved \\
anti-rem & 27.7$\pm$7.9 & 13.1 & 2.11$\pm$0.60 & VDM+direct \\
\hline
\hline
\end{tabular}
\caption{
Total cross section (pb) of $\Lambda (\overline{\Lambda})$ in the
$|\cos \theta |<0.77$ and $0.75<P_T<2.75$ GeV range.
The notation (-VDM) means the VDM subtraction which was described in the
text.
Here, we use the LO theories in order to show the discrepancy
with the experimental data.
}
\label{ttotal}
\end{table}
In order to extract the NLO effect, we calculated
the LO cross sections, which are also shown in Table \ref{ttotal}.
Here, we removed the NLO factorization from our Monte-Carlo generator.
The ratio between the theoretical
predictions and the experimental data are typically $\sim$2.

\section{Conclusion}
We carried out an inclusive measurement of
$\Lambda(\overline{\Lambda})$ productions
in two-photon
processes at TRISTAN. The mean $\sqrt{s}$ was 58 GeV and the
integrated luminosity was 265 pb$^{-1}$.
Inclusive $\Lambda (\overline{\Lambda})$ cross sections
in two-photon processes
were obtained under such conditions as
no-electron, anti-electron, and remnant-jet tags.
In particular, using the remnant-jet tagging
we could unambiguously extract the contribution
from the resolved-photon process.
Comparisons with theoretical predictions were carried out.
The cross sections were two-times larger than the theoretical predictions
without hard gluon emissions, suggesting the necessity of
next-to-leading-order Monte-Carlo generator.

\section*{Acknowledgement}

We thank Drs. K. Hagiwara and T. Sj\"ostrand for helpful discussion
concerning the baryon production mechanism.
We also thank the TRISTAN accelerator
staff for the successful operation of TRISTAN.
The authors appreciate
all of the engineers and technicians at KEK as well as
those of the collaborating
institutions: H. Inoue, N. Kimura, K. Shiino, M. Tanaka, K. Tsukada, N. Ujiie,
and H. Yamaoka.

\end{document}